\documentclass[preprint2]{aastex}
\usepackage{amsmath}

\shorttitle{Stability analysis for power-law spherical accretion}
\shortauthors{X. Hernandez, L. Nasser and P. L. Rendon}

\begin{document}


\title{Formal stability analysis for the recent $\gamma=5/3$ power-law spherical accretion solution}


\author{X. Hernandez$^{1}$, L. Nasser$^{2}$ and P. L. Rend\'on$^{3}$}
\affil{$^{1}$Instituto de Astronom\'{\i}a, Universidad Nacional Aut\'{o}noma de M\'{e}xico, Apartado Postal 70--264, 
04510, Mexico City, Mexico.\\
$^{2}$ Department of Science and Mathematics, Columbia College, Chicago, IL 60605, USA. \\
$^{3}$ Instituto de Ciencias Aplicadas y Tecnología, Universidad Nacional Autónoma de México, Circuito Exterior S/N, Ciudad
Universitaria, 04510, Mexico City, Mexico.\\ }



\begin{abstract}

  Recently, an exact spherically symmetric analytic accretion solution was presented having simple $\rho \propto R^{-3/2}$ and
  $V \propto R^{-1/2}$ scalings in Hernandez et al. (2023). In dimensionless variables that solution forms a one-parameter
  family of solutions ranging from an empty free-fall solution to a hydrostatic equilibrium configuration. This power-law solution
  is characterised by a constant Mach number for the flow, which can vary from zero to infinity as a function of the one parameter
  of the solution, and has an accretion density profile which naturally goes to zero at large radii. This accretion density profile
  was shown in Hernandez et al. (2023) to be an accurate representation of the accretion density profiles of a sample of AGN galaxies,
  over hundreds of Bondi radii. The observed density profiles fall by many orders of magnitude in density beyond their Bondi radii,
  something which is inconsistent with classical Bondi models where the accretion density profiles rapidly converge to a constant
  outside of the Bondi radius. While the good agreement with observations is suggestive of a global stability for the solution mentioned,
  no formal stability analysis for it has previously been presented. Here we perform such stability analysis and show the solution
  mentioned to be globally stable for all values of the parameters governing it, both for its accretion and outflow modes. This
  result makes the $\gamma=5/3$ power-law spherical accretion model an interesting analytical addition to the study and description of
  accretion problems in astrophysics.

\end{abstract}



\section{Introduction}

In the context of astrophysical accretion problems, the case of spherical accretion is generally treated through
the Bondi (1952) model, which yields general purpose relations between accretion rates and typical density and
sound speed parameters for gas accreting onto a central object of a given mass. To mention but a few recent examples,
Raghuvanshi \& Dutta (2023) in the modelling of the formation of Population III stars, estimations of the growth
rate of early galactic black holes by Trinca et al. (2023) or even the accretion of hypothetical axion
dark matter onto primordial black holes in Mazde \& Visinelli (2023).

Many of the systems to which the Bondi model is routinely applied however, are clearly inconsistent with the structure
of the accretion density profiles implicit to the Bondi solution. In  particular, the Bondi model assumes a
density for the accreting material which tends to a finite value at infinity. Indeed, for any maximum accretion
Bondi model of the type generally used to derive accretion rate scalings, the density of the accreting material
quickly converges to its large radius asymptote on crossing the Bondi radius. Observations in contrast, e.g. the de-projected
AGN accretion density profiles of Pl\v{s}ek et al. (2022), show negative power-law scalings which continue to fall
by many orders of magnitude after crossing the inferred Bondi radii of the galaxies studied.

With this concern in mind, two of us recently presented in Hernandez et al. (2023) a new exact spherical
accretion solution for $\gamma=5/3$ where the density profile of the accreting material is a simple $R^{-3/2}$
power-law. The accreting density of this model hence tends to zero for large radii, as happens with the
velocity profile, which is given by a $R^{-1/2}$ scaling. In that paper a number of interesting results regarding
the non-spherical perturbations of the model were presented, yielding equatorial infall and polar outflow
configurations, through purely hydrodynamical mechanisms, consistent with the results of recent detailed
numerical experiments from our group and others e.g., Aguayo-Ortiz et al. (2019), Waters et al. (2020), Tejeda et al. (2020).

In Hernandez et al. (2023), the relevance of the model was shown through comparing it to de-projected accretion
density profiles from x-ray observations of various AGN galaxies from Allen et al. (2006), Runge \& Walker (2021) and Pl\v{s}ek et al. (2022).
The comparison was in many cases excellent, and even in cases where the fit was poor, it remained much better than that of a Bondi model, where
the accretion density profile flattens outwards of the Bondi radius, for the AGN cases treated where the accretion
density profiles continue to fall by many orders of magnitude in density for the entire observed regions spanning
hundreds of Bondi radii.

The temporal stability of the new solution mentioned however, was merely suspected on account of the good
agreement with observations, and on the basis of a limited number of preliminary numerical experiments.
In this short paper we present a full formal stability analysis of the exact power-law $\gamma=5/3$ accretion
model of Hernandez et al. (2023).

This paper is organised as follows: Section 2 briefly presents the new exact spherical steady-state
hydrodynamic power-law solution of Hernandez et al. (2023), in the interest of making the present paper
self-consistent and as an introduction to the notation to be employed. Section 3 then develops a formal stability
analysis of the steady-state solution in question through the standard inclusion of the time-dependent terms in
the structural equations of the problem and the assumption of a general solution consisting of the steady-state
solution plus the addition of a small time-dependent perturbation. The conservation equations are then linearized
with respect to the perturbation. Solving the resulting system of equations proves the stability of the power-law
solution of Hernandez et al. (2023), as no instability regime appears. Dispersion relations and perturbation velocities
are also given for the high frequency limit. Concluding statements appear in Section 4.

\section{Spherically symmetric $\gamma=5/3$ power-law accretion solution} 

We now present the exact power-law solution for the spherical accretion problem of Hernandez et al. (2023) and a brief summary of its main
features. Although the material in this section already appears in Hernandez et al. (2023), it is included here for context, self-consistency
of the present paper and to introduce the notation used. We begin from the equations of mass and radial momentum conservation in spherical
coordinates:

\begin{equation}
\frac{\partial \rho}{\partial t}  +\frac{1}{r^{2}} \frac{\partial(r^{2} \rho V)}{\partial r} =0,
\end{equation}

\begin{equation}
\frac{\partial V}{\partial t} + V\frac{\partial V}{\partial r} = -\frac{1}{\rho}\frac{\partial P}{\partial r} - \frac{G M}{r^{2}},
\end{equation}

\noindent where $\rho(r,t)$, $P(r,t)$ and $V(r,t)$ are the gas density and pressure, and the $r$  velocity, respectively e.g.
Binney \& Tremaine (1987). Assuming a barotropic equation of state $P=K \rho^{\gamma}$ eq.(2) yields:

\begin{equation}
\frac{\partial V}{\partial t}  +V\frac{\partial V}{\partial r}  = -K \gamma \rho^{\gamma-2} \frac{\partial \rho}{\partial r} - \frac{G M}{r^{2}},
\end{equation}

Now we introduce dimensionless variables (e.g. Bondi 1952) $\varrho=\rho/\bar{\rho}$, $\mathcal{V}=V/\bar{c}$, $R=r/\bar{r}$ and $\tau=t/ \bar{t}$
where $\bar{\rho}$ is a reference density at a certain point, $\bar{c}^{2}=K \gamma \bar{\rho}^{\gamma-1}$, the sound speed at this same reference point,
$\bar{r}=GM/\bar{c}^{2}$ and $\bar{t}=\bar{r}/\bar{c}$. Equations (1) and (3) now read:

\begin{equation}
\frac{\partial \varrho}{\partial \tau}    +\frac{1}{R^{2}} \frac{\partial(R^{2} \varrho \mathcal{V})}{\partial R} = 0,
\end{equation}

\begin{equation}
  \frac{\partial \mathcal{V}}{\partial \tau} + \mathcal{V}\frac{\partial \mathcal{V}}{\partial R} =
  -\varrho^{\gamma-2} \frac{\partial \varrho}{\partial R} - \frac{1}{R^{2}}.
\end{equation}

\noindent Equations (4) and (5) define the problem. Firstly we turn to a simple steady-state power-law solution valid for $\gamma=5/3$:

\begin{equation}
\mathcal{V}=\mathcal{V}_{0}R^{-1/2},
\end{equation}

\begin{equation}
\varrho=\varrho_{0}R^{-3/2},  
\end{equation}

\noindent where $\mathcal{V}_{0}$ and $\varrho_{0}$ are two constants satisfying eq.(5):

\begin{equation}
\mathcal{V}_{0}^{2}=2-3\varrho_{0}^{2/3}.  
\end{equation}

\noindent The boundary conditions for this solution are fixed at infinity where the fluid is at rest with zero density and hence zero sound speed.
For this solution equation (4) is readily integrated to give the dimensionless mass accretion rate as:

\begin{equation}
\dot{M}=4 \pi \mathcal{V}_{0}\varrho_{0}.
\end{equation}

\noindent Equation (8) implies that the solution will exist only for the $0<\varrho_{0}^{2/3}<2/3$ interval, where the $\mathcal{V}_{0}$ constant
will be restricted to the interval $2>\mathcal{V}_{0}^{2}>0$. When modelling an infall solution $\mathcal{V}_{0}$ will be negative, and positive
if one is modelling an outflow solution. Towards the  $\varrho_{0} \to 0$ limit the pressure vanishes, hydrodynamical effects disappear and we obtain
an empty state in free-fall at $\mathcal{V}_{0}=-\sqrt{2}$ for the infall solution and expanding at $\mathcal{V}_{0}=\sqrt{2}$ for the outflow one, while the
$\varrho_{0}^{2/3} \to 2/3$ limit is a hydrostatic equilibrium state. While the accretion rate vanishes at these two limits, a maximum appears at an
intermediary value of  $\mathcal{V}_{0}^{2}$ (or alternatively $\varrho_{0}^{2/3}$), showing the steady-state power-law solution to have only one free
parameter for a description in dimensionless variables.

Equation (8) can be used to write eq.(9) in terms of $\varrho_{0}^{2/3}$ and $\dot{M}$ only:

\begin{equation}
\dot{M}^{2}=(4 \pi \varrho_{0}^{2})^{2}\left(2- 3\varrho_{0}^{2/3}  \right).
\end{equation}

\noindent The maximum accretion rate is now obtained by differentiating the above equation with respect to $\varrho_{0}$ and equating to zero.
The maximum accretion rate occurs at $\varrho_{0c}^{2/3}=\mathcal{V}_{0c}^{2}=1/2$ and $\dot{M}=\pi$.


To conclude this section, we calculate the Mach number of the flow which follows from dividing the velocity in eq.(6) by the local sound
speed, which in dimensionless variables is given by $\varrho^{1/3}$, resulting in:

\begin{equation}
\mathcal{M}=\mathcal{V}_{0}\varrho_{0}^{-1/3}.
\end{equation}

From the above equation we see the unexpected result of $\mathcal{M}=1$ at all radii for the maximum accretion rate case of
$\varrho_{0}^{2/3}=\mathcal{V}_{0}^{2}=1/2$. The lack of a radial dependence for the Mach number is general and applies also towards
the hydrostatic equilibrium  $\varrho_{0}^{2/3}=2/3$, $\mathcal{M}=0$ limit, where the flow becomes sub-sonic in the $1/2 < \varrho_{0}^{2/3} < 2/3 $  interval.
Within the  $0 < \varrho_{0}^{2/3} < 1/2 $ interval, $\mathcal{M}>1$ and diverges towards the  $\varrho_{0}=0$ free-fall limit where the sound
speed becomes zero.


Contrary to the general result of the Bondi model where a sonic radius always appears, the situation in the power-law model is distinct and
characterised by a radially constant Mach number, at a value given by the single parameter of the solution. Therefore, no shocks develop in the
power-law solution as the flow is never trans-sonic. This power-law solution is thermodynamically more limited than the Bondi solution, as the
former is valid only for $\gamma=5/3$, while the latter is valid for all $\gamma \leq 5/3$. At this last value the Bondi model has a singular point,
as in that case the sonic radius goes to the origin. 

This completes the summary of the $\gamma=5/3$ power-law solution, first presented and explored more extensively in Hernandez et al. (2023).
In that paper two of us also presented interesting results regarding the perturbative inclusion of angular momentum, small variations about
$\gamma=5/3$ and non-spherical perturbations for this power-law solution. The temporal stability of the model given by eqs.(6)-(8) was suggested
in that publication from the excellent agreement of the density profile of this model across several orders of magnitude in radius to the
observationally inferred de-projected accretion density profiles for a sample of AGN observations from Pl\v{s}ek et al. (2022), and from the
stability of a small number of numerical experiments. A formal stability analysis of the power-law accretion model of eqs.(6)-(8) is developed
in the following section.

\section{Stability analysis for the spherically symmetric $\gamma=5/3$ power-law accretion solution}

We shall follow the standard approach of introducing a small time-dependent perturbation on our steady-state solution and working through with the
resulting linearized system, e.g. Bender \& Orszag (1978). We begin by considering a solution to the system of equations (4) and (5) of the form:

\begin{equation}
\varrho(R,\tau)=\varrho_{0}R^{-3/2}+\epsilon \varrho_{1}(R,\tau),
\end{equation}

\begin{equation}
\mathcal{V}(R,\tau)=\mathcal{V}_{0}R^{-1/2}+\epsilon \mathcal{V}_{1}(R,\tau),
\end{equation}

\noindent a perturbative time dependence added onto the steady-state solution of eqs.(6) and (7), where we shall assume $\epsilon<<1$. Notice that
the condition $\mathcal{V}_{0}^{2}=2-3\varrho_{0}^{2/3}$ will still apply. Introducing the above into equations (4) and (5) and keeping only terms to
first order in $\epsilon$ yields:

\begin{equation}
\begin{split}
  R^{3/2}\frac{d \varrho_{1}}{d \tau} +\mathcal{V}_{0} \left( R\frac{d \varrho_{1}}{d R} + \frac{3 \varrho_{1}}{2} \right)\\
  +\frac{\varrho_{0}}{R} \left( R\frac{d \mathcal{V}_{1}}{d R} + \frac{ \mathcal{V}_{1}}{2} \right)=0,
\end{split}
\end{equation}

\begin{equation}
\begin{split}
  R^{1/2}\frac{d \mathcal{V}_{1}}{d \tau} + \varrho_{0}^{-1/3} \left( R\frac{d \varrho_{1}}{d R} + \frac{\varrho_{1}}{2} \right)\\
  +\frac{\mathcal{V}_{0}}{R}  \left( R\frac{d \mathcal{V}_{1}}{d R} - \frac{ \mathcal{V}_{R}}{2} \right)=0.
\end{split}
\end{equation}

Now we introduce a separation of variables ansatz with the temporal dependence of both the perturbed velocity and density modelled as
a complex exponential,

\begin{equation}
\varrho_{1}=\varrho_{r}(R) e^{i \omega \tau}, \,\,\,\,\,\,\,\,\,\,  \mathcal{V}_{1}=\mathcal{V}_{R}(R) e^{i \omega \tau},
\end{equation}

\noindent and switch to a description of the relative perturbations normalised by the steady-state solution through the change of variables:

\begin{equation}
\Delta_{\varrho}(R)=\frac{\varrho_{R}}{\varrho_{0}}R^{3/2},  \,\,\,\,\,\,\,\,\,\, \Delta_{\mathcal{V}}(R)=\frac{\mathcal{V}_{R}}{\mathcal{V}_{0}}R^{1/2},
\end{equation}

\noindent as used in the classical study of stability of stellar interiors towards
radial modes e.g. Hansen \& Kawaler (1994). Equations (14), (15) now read:

\begin{equation}
    \frac{i \omega}{\mathcal{V}_{0}} R^{1/2} \Delta_{\varrho}+\frac{d \Delta_{\varrho}}{dR} +\frac{d \Delta_{\mathcal{V}}}{dR}=0,
\end{equation}

\begin{equation}
\begin{split}
  i \omega \mathcal{V}_{0} R^{1/2} \Delta_{\mathcal{V}}+\varrho_{0}^{2/3}\left( \frac{d \Delta_{\varrho}}{dR} -\frac{\Delta_{\varrho}}{R} \right)\\
  +\mathcal{V}_{0}^{2}\left( \frac{d \Delta_{\mathcal{V}}}{dR}-\frac{\Delta_{\mathcal{V}}}{R}\right)=0.
\end{split}
\end{equation}

At this point we introduce a radial re-scaling $\mathcal{R}=R^{3/2}$ and a further change of variables:

\begin{equation}
  X_{\mathcal{R}}(\mathcal{R})=\Delta_{\varrho}+\Delta_{\mathcal{V}} \,\,\,\,\,  Y_{\mathcal{R}}(\mathcal{R})=\varrho_{0}^{2/3}\Delta_{\varrho}+
  \mathcal{V}_{0}^{2}\Delta_{\mathcal{V}} 
\end{equation}

Which turns eqs. (18) and (19) into:

\begin{equation}
  i \omega Y_{\mathcal{R}}-i \omega \mathcal{V}_{0}^{2} X_{\mathcal{R}}+ (1-2 \mathcal{V}_{0}^{2}) \mathcal{V}_{0}  \frac{d X_{\mathcal{R}}}{d \mathcal{R}}=0,
\end{equation}

\begin{equation}
\begin{split}
  i \omega \mathcal{V}_{0} \mathcal{R} (2-\mathcal{V}_{0}^{2})X_{\mathcal{R}} -3i\omega \mathcal{V}_{0}\mathcal{R}Y_{\mathcal{R}}\\
  +3(1-2\mathcal{V}_{0}^{2}) \mathcal{R} \frac{d Y_{\mathcal{R}}}{d\mathcal{R}}-2(1-2\mathcal{V}_{0}^{2} )Y_{\mathcal{R}}=0,
\end{split}
\end{equation}

\noindent where we have used condition (8) to eliminate $\varrho_{0}^{2/3}$ in favour of $\mathcal{V}_{0}^{2}$. We can now solve for $Y_{\mathcal{R}}$
from eq.(21) and introduce it into eq.(22) to yield an equation in $X_{\mathcal{R}}$ alone, 

\begin{equation}
\begin{split}
  3\mathcal{R}(2\mathcal{V}_{0}^{2}-1)\frac{d^{2}X_{\mathcal{R}}}{d\mathcal{R}^{2}}+6i\omega\mathcal{V}_{0}\mathcal{R}\frac{d X_{\mathcal{R}}}{d\mathcal{R}}
  -2\omega^{2}\mathcal{R}X_{\mathcal{R}}\\
  +2(1-2\mathcal{V}_{0}^{2})\frac{d X_{\mathcal{R}}}{d\mathcal{R}}-2i\omega\mathcal{V}_{0} X_{\mathcal{R}} =0.
\end{split}
\end{equation}

\noindent The above has as solution:

\begin{equation}
\begin{split}
X_{\mathcal{R}}=X_{1} \mathcal{R}^{5/6} e^{\left(\frac{i \omega \mathcal{V}_{0}\mathcal{R} }{1-2 \mathcal{V}_{0}^{2}}\right)} J_{5/6}\left(\frac{\omega \mathcal{R}
  [2- \mathcal{V}_{0}^{2}]^{1/2} }{\sqrt{3} \left[1- 2 \mathcal{V}_{0}^{2} \right]}\right)\\
 +X_{2} \mathcal{R}^{5/6} e^{\left(\frac{i \omega \mathcal{V}_{0}\mathcal{R} }{1-2 \mathcal{V}_{0}^{2}}\right) }Y_{5/6}\left(\frac{\omega \mathcal{R}
  [2- \mathcal{V}_{0}^{2}]^{1/2} }{\sqrt{3} \left[1- 2 \mathcal{V}_{0}^{2}\right]}\right),
\end{split}
\end{equation}

\noindent where $J_{n}(x)$ and $Y_{n}(x)$ are the quasi-periodic Bessel functions of the first and second kind, respectively, and $X_{1}$ and $X_{2}$
are two amplitude constants. The real part of the above equation gives the perturbation in $X$ in the re-scaled variables, for given values of $\omega$
and of the $\mathcal{V}_{0}$ parameter of the steady-state spherical power-law solution, with $X_{1}$ and $X_{2}$ determined by the initial conditions of
the perturbation chosen. As is customary (e.g. Rienstra 1999) we take $X_{2}=0$ on account of the divergence of $Y_{5/6}(x)$ as $x \to 0$, which would
clearly invalidate the perturbative ansatz within which we are working.

The first thing to notice is that within any bound radial interval, $X$ is given by bound oscillatory functions, proving the global stability of the
power-law $\gamma=5/3$ solution of eqs. (6)-(8) to small perturbations of any frequency since the argument of $J_{5/6}$ never becomes imaginary as
$0<\mathcal{V}_{0}^{2}<2$. There are no instability scales for the problem either at the accretion or the outflow regimes, either at the sub-sonic
or the super-sonic ones.

We see also that since as $x \to \infty$ $J_{5/6}(x) \to -(2/\pi x)^{1/2} \sin(\pi/6-x)$, as $\mathcal{R} \to \infty$,  $X$ scales with $\mathcal{R}^{1/3}$.
Therefore, $X$ diverges as $\mathcal{R} \to \infty$. This imposes a large scale maximum validity range for the solution and the stability criterion developed,
through requiring consistency with the perturbative approach used. This large scale limit however, can be taken as large as desired by taking an adequately
small value of $X_{1}$. In any case, in any practical application a finite outer radius for the system in question will always appear, e.g. the
galactic radius for the AGN accretion density profiles reported by Pl\v{s}ek et al. (2022) and modelled using eq. (7) in Hernandez et al. (2023).

{ 
We now turn to the high frequency limit e.g. Papaloizou \& Pringle (1984), where for the limit of $J_{5/6}(x)$ as $x\to \infty$ given above,
$X$ will tend to:

\begin{equation}
\begin{split}
  X_{1} \left( \frac{2 \sqrt{3} \left[1- 2 \mathcal{V}_{0}^{2} \right] }{\pi \omega [2- \mathcal{V}_{0}^{2}]^{1/2} } \right)^{1/2} \mathcal{R}^{1/3}
  \cos \left( \frac{\omega \mathcal{V}_{0} \mathcal{R}}{1-2\mathcal{V}_{0}^{2}} \right) \\
  \times \sin \left(\frac{\pi}{6}-\frac{\omega \mathcal{R}[2- \mathcal{V}_{0}^{2}]^{1/2} }{\sqrt{3} \left[1- 2 \mathcal{V}_{0}^{2} \right]} \right).
\end{split}
\end{equation}

\noindent Using now the identity $2\cos(A)\sin(B)=\sin(A+B)-\sin(A-B)$ we see that the high frequency limit for $X$ will be
characterised by the sum of two sine functions having the two wavenumbers $k_{\pm}$ given below:

\begin{equation}
  k_{\pm}=\omega \left( \frac{\sqrt{3}\mathcal{V}_{0} \pm [2- \mathcal{V}_{0}^{2}]^{1/2}}{\sqrt{3}[1-2 \mathcal{V}_{0}^{2}]}   \right), 
\end{equation}

%

\noindent which becomes the dispersion relation for the problem in the high frequency limit.} In this limit the reaction of the system to perturbation of
wavenumbers $k_{\pm}$ will be of temporal oscillations at a frequency given by the value of $\omega$ in equation (26), for a given value of the $\mathcal{V}_{0}$
parameter of the unperturbed power-law solution of eqs.(6)-(8). Firstly, notice from equation (21) that the perturbation for $Y_{\mathcal{R}}$ will also be of
periodic character with the same wavelength as that of a periodic $X_{\mathcal{R}}$.

Given that for the unperturbed solution $0<\mathcal{V}_{0}^{2}<2$, it is important to note that the argument of the square root appearing in eq.(26) will
always be $>0$, and hence $\omega$ never picks up an imaginary component for any real value of $k_{\pm}$. Therefore, the reaction of the system in the high
frequency limit to perturbations of any wavelength, $\lambda_{\pm} =2\pi/k_{\pm}$, will always be oscillatory in time with a frequency given by
$\omega$ in eq.(26).



We can now calculate the two propagation velocities of perturbations in the high frequency regime, for any wavelength $k_{\pm}$, as
$c_{\pm}=-2 \pi \omega/k_{\pm}$, where the leading minus sign in the preceding expression comes from the way the temporal perturbations
were defined in eq. (16), to yield:

\begin{equation}
c_{\pm}=\frac{2\pi\sqrt{3}(2 \mathcal{V}_{0}^{2}-1) }{\sqrt{3}\mathcal{V}_{0} \pm (2- \mathcal{V}_{0}^{2})^{1/2} }.
\end{equation}

\noindent We see that the problem is non-dispersive, with the propagation velocity of perturbations depending on the $\mathcal{V}_{0}$ parameter of
the unperturbed solution, but never becoming a function of the wavelength of the perturbations.

\begin{figure}[t]
\hskip -10pt \includegraphics[width=8.0cm, height=7.5cm]{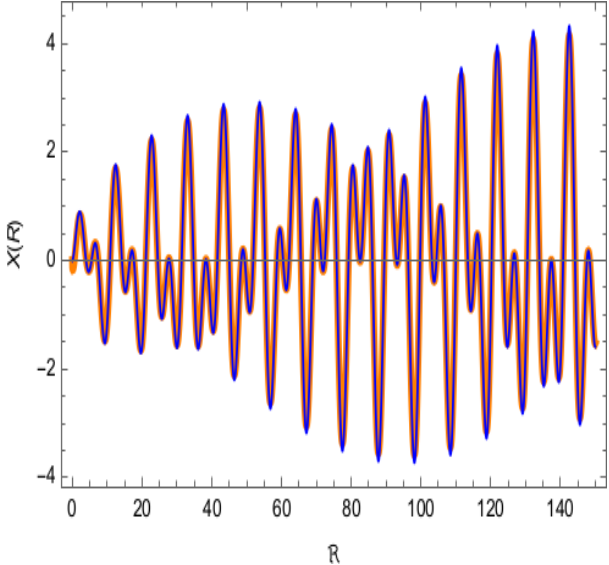}
\caption{Exact solution to eq.(24) for $X_{1}=1$, $X_{2}=0$, $\omega=1$ and $\mathcal{V}_{0}=-1/4$, thin dark curve, superimposed on the high
  frequency limit of the solution from eq.(25), thick light curve.}
\end{figure}

The perturbation propagation velocities of eq.(27) for the accretion mode of $\mathcal{V}_{0}<0$ are always inward for the plus sign, and change
from being inward in the super-sonic $2>\mathcal{V}_{0}^{2}>1/2$ regime to outward propagation in the sub-sonic $1/2>\mathcal{V}_{0}^{2}>0$
regime. For the outflow mode of the unperturbed solution of $\mathcal{V}_{0}>0$, the minus sign in eq.(26) results in outward propagating
perturbations for all values of $\mathcal{V}_{0}$, but for the plus sign we obtain outward propagation for the super-sonic regime and inward
for the sub-sonic one. The cases where one of the propagation velocities of the perturbations change sign are separated by a zero velocity at
$\mathcal{V}_{0}=1/2$, $\mathcal{M}=1$, for one of the two waves appearing, which becomes static with only the other mode propagating. Still,
in keeping with the overall stability of the solution, these non-propagating solutions do not grow. We hence see that going from the accretion
$ \mathcal{V}_{0}<0$ to the radial outflow $ \mathcal{V}_{0}>0$ mode of the power-law solution has no effect on the stability of the solution,
the only difference being a switch in the propagation direction of some of the perturbations, as described above.

The point $\mathcal{R}=0$ lies outside of the high frequency limit, as the argument of $J_{5/6}$ in eq.(24) goes to zero and
we obtain $X=0$, fixing the boundary conditions of the perturbations being modelled as $X(\mathcal{R}=0)= 0$. This does not imply
that the perturbations in density and velocity vanish at the origin, only that the relative perturbations cancel each other out, from the definition
of the variable $X$ in eq. (20).

Notice that the high frequency limit is approached quite rapidly. This is illustrated in figure 1 where we show the exact solution of eq.(24)
for $X_{1}=1$, $X_{2}=0$, $\omega=1$ and $\mathcal{V}_{0}=-1/4$, thin dark curve, superimposed on the high frequency limit of the solution for the same parameters,
eq(25), thick light curve. The double wavenumber structure is evident, as is the very good agreement between the exact solution and the high frequency limit;
both curves appear practically on top of each other almost from the start. Other particular examples yield equivalent results.

\section{Conclusions}

For the recent exact $\gamma=5/3$ accretion model of Hernandez et al. (2023) we have performed a rigorous stability analysis and shown the
model to be stable in all of its regimes, as no instability scales arise neither in the accretion nor outflow mode of the model, for any
value of the $\mathcal{V}_{0}$ parameter which determines the model. This global stability of the exact accretion solution highlights
its relevance as a modelling tool for real astrophysical accretion or outflow problems.

In the high frequency limit an analytic dispersion relation is obtained, with spatial perturbations which become periodic with well defined pairs
of wavenumbers and propagation velocities. The resulting dispersion relations are non-dispersive. Propagation velocities for the perturbations in the high
frequency limit can be of either inward or outward propagation, depending on the details of the parameters chosen.

\section*{Acknowledgements}
Xavier Hernandez acknowledges financial assistance from UNAM DGAPA PAPIIT grant IN106220 and CONACYT. L. Nasser gratefully acknowledges the
support from the NSF award PHY - 2110425. Pablo L. Rendon acknowledges financial assistance from UNAM DGAPA PAPIIT grant IN117823.

\end{document}